# State Space Reduction
# For Reachability Graph of CSM Automata

by

Wiktor B. Daszczuk [1]

**Institute of Computer Science**
**Warsaw University of Technology**

Nowowiejska Str. 15/19, 00-665 Warsaw, Poland

Summary

Classical CTL temporal logics are built over systems with interleaving model concurrency. Many attempts are made to fight a state space explosion problem (for instance, compositional model checking). There are some methods of reduction of a state space based on independence of actions. However, in CSM model, which is based on coincidences rather than on interleaving, independence of actions cannot be defined. Therefore a state space reduction basing on identical temporal consequences rather than on independence of action is proposed. The new reduction is not as good as for interleaving systems, because all successors of a state (in depth of two levels) must be obtained before a reduction may be applied. This leads to reduction of space required for representation of a state space, but not in time of state space construction. Yet much savings may occur in regular state spaces for CSM systems.

[1] wbd@ii.pw.edu.pl

# I. Introduction

For efficient evaluation of temporal formulas in CTL temporal logic [BenA83, Bose90, Burc90, Clar86, Clar89, Clar92, Lamp83, Mann81, Mann84, Pnue77, Pnue85, Wolp86], an algorithm of finding smallest or largest fixed point of special functional over a reachability graph of a system is often applied [McMi92]. If a ROBDD (Reduced Ordered Binary Decision Diagrams) representation is used [Brya86], it is called a symbolic model checking [McMi92]. Every component of a system is assigned an arbitrary set of boolean variables (for example, $\{x_1, x_2\}$ for a component having 3 or 4 states), and every state of a component is represented as a boolean function over this set ($x_1 \wedge \neg x_2$, for example). States of state space are simply conjunctions of functions representing states of components. If a state $s_1$ of components $c_1$ has a representation $x_1 \wedge \neg x_2$, and a state $r_1$ of component $c_2$ has a representation $y_1 \wedge y_2$, then a state being a superposition of these two states $(s_1, r_1)$ in state space of the two-component system has a boolean function $(x_1 \wedge \neg x_2) \wedge (y_1 \wedge y_2)$ as a representation.

ROBDD representation allows to represent large state spaces in compact form, in general consuming much less memory than exact representation [McMi92]. Moreover, the algorithms of evaluation of temporal formulas using ROBDD are very simple and efficient, because they exploit the feature that sets of states and set operations like union or intersection are represented naturally as boolean functions [Dasz00b].

- set is represented as disjunction of functions representing its elements;
- union is represented as disjunction of functions representing its arguments;
- intersection is represented as conjunction of functions representing its arguments;
- complement is represented as negation of function representing its argument.

Although a ROBDD representation is very compact and efficient, it is sometimes still too large to be stored in computer memory and to evaluate temporal formulas on it. Therefore, many attempts were made to obtain a reduced state space for temporal verification. All of them work on interleaving models of computations. In the paper a new algorithm is presented for CSM systems [Mieś92a, Mieś92b, Mieś94], which is constructed over coincidences of action rather than on interleaving.

In section II, a system of CSM automata is presented together with its state space – Reachability Graph [Mieś92b]. In section III a QsCTL temporal logic is constructed over Reachability Graph of CSM automata. In section IV, an invisibility of actions is defined, which is required for reduction. Section V contains the reduction rule in CSM. For the purpose of the reduction, some additional atomic formulas must be added to the basic set. The principle of acquiring a complete set of atomic boolean formulas is presented in section VI. A proof of correctness of the reduction rule is given in section VII. A shape of reachability graph after reductions applied is described in section VII. Section IX contains an algorithm of reduction. Time complexity of the algorithm is analyzed in section X. Conclusions are presented in section XI.

# II. Reachability Graph of a system of CSM automata.

A system of CSM automata is presented in [Mieś92a, Mieś92b, Mieś94]. The general features of CSM automata are:

- every component automaton has exactly one initial state;
- component automata are Moore-like (signals are generated in states);

- arcs are labeled with boolean formulas over input alphabet (if formula is fulfilled, the arc may be followed, for example if signal *q* is active then the formula $q \vee p \wedge m$ is fulfilled);
- arcs leading form a state to the same state are allowed (they are called **ears**);
- automata are complete, i.e. the disjunction of all formulas on arcs leading out of a given state equals *true*;
- signals (letters of input alphabet) come from output of automata and from external world;
- input and output alphabets are not disjoint (an automaton generates signals for himself of for other automata);
- special symbols denote the formulas: $\mathbb{1}$-always true (an arc may be always followed), $\mathbb{0}$-always false (a lack of arc);
- if more than one formula is fulfilled on arcs leading out of a given state – the transitions is chosen in non-deterministic way;

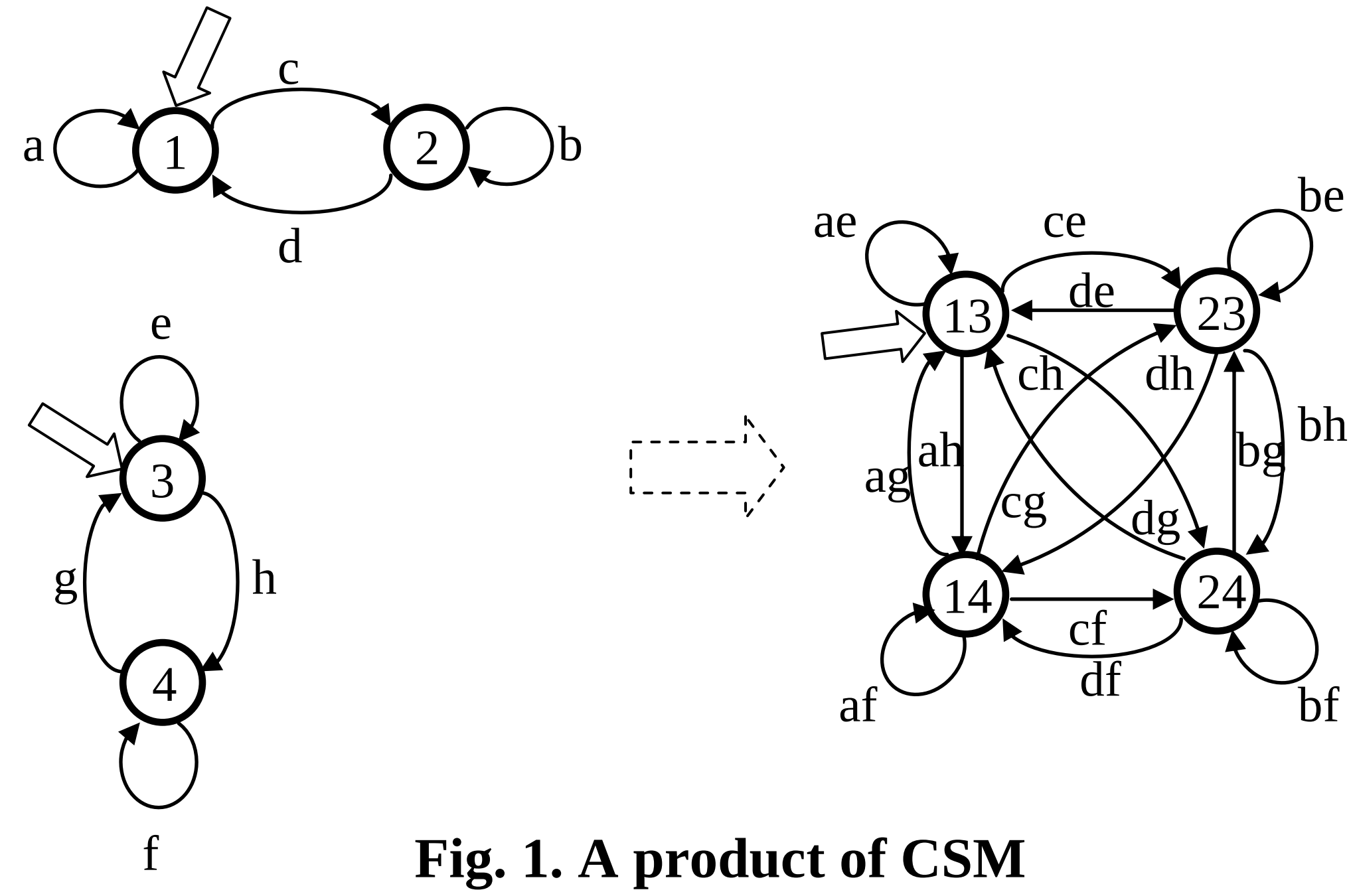

**Fig. 1. A product of CSM Automata - RG**

- all automata in a system perform always one transition synchronously in a lock-step manner (no external clock is required)

The state space of a system of CSM automata, called Reachability Graph (RG), is obtained as presented in Fig. 1. States of RG are superpositions of states of component automata. Arcs between states are obtained as products of arcs of component automata, i.e. formula on arc of RG is a conjunction of formulas on arcs in component automata (letters on arcs in Fig. 1. represent boolean formulas rather than single letters of input alphabet).

A set of signals generated in a state of RG is a unions of sets of signals generated in states of components automata. For simplicity, output signals generated in states are omit in Fig. 1. If the state *1* generate signals *p,q*, and state *3* generates signals *q,m*, then the state *13* of RG generates signals *p,q,m*.

The complete algorithm of obtaining RG from component CSM automata is given in [Mieś92b].

## III. The Temporal Logic

The QsCTL temporal logic constructed over RG is as follows:

- The set of states of Kripke Structure is simply a set of states of Reachability Graph of CSM automata.
- The succession relation is simply a set of arcs of RG (excluding ears leading out of non-terminal states, therefore a graph is denoted $RG_{-@}$).
- The initial state is an initial state of RG.
- The set of atomic boolean formulas $\pi$ is:
  - ⇒ a signal being generated in states of RG (denoted as a name of the signal),
  - ⇒ staying in a given state of RG (denoted ***in*** *s*),
  - ⇒ staying in one of a set *S* of states of RG (denoted ***in*** *S*),
  - ⇒ staying in a state *s* of RG having a as projection on a component automaton ***a*** given state $s_{\mathbf{a}}$.

  A set of atomic boolean formulas *true* in a given state *s* is denoted $\pi(s)$.

The modalities used in the temporal logic are following:

- □ - always (often denoted AG),
- ◊ - eventually (often denoted AF),
- ○ - next (often denoted AX),
- **U** - until (weak until, often denoted $AU_W$),
- $\circ_{\mathbf{a}}$ - next in automaton **a**, true by definition in a state having a as projection on an automaton **a** a terminal state.

Additionally, a quantified formulas of a form $\boldsymbol{Q}s \in S;\ s\!:\varphi$ (***Q*** is $\forall$ or $\exists$) are allowed. Such formulas are not allowed when using classical algorithm of evaluation. In classical algorithm, an evaluation is performed bottom-up (sets of states where atomic formulas are fulfilled are evaluated first, and then embracing formulas). The formula ***in*** *s*, where *s* is a state variable passing through a scope of a quantifier, cannot be evaluated [Dasz00a].

Checking in spheres (see section V) is performed top-down and therefore quantified formulas may be evaluated.

For definition of QsCTL temporal logic for systems of CSM automata see [Dasz].

## IV. Invisibility of states

The obtaining of $RG_{-@}$ for a set of CSM automata in general leads to the explosion of states (a Cartesian product of all states of component automata in worst case). The manner of reduction of a state space proposed for interleaving system in [Gert99] cannot be applied to CSM systems, because no pair of states in component automata can be expected to be independent for sure. For example, an automaton may omit an action if it is informed (via signal) that another automaton has performed the action in past.

Instead, I propose a reduction that is based on deleting arcs if it does not change temporal consequences. More precisely, the reduction is based on deleting invisible arcs (see later for definition 1 of invisibility) from RG$_{—@}$. If an arc leading from $s_i$ to $s_j$, then every pair of arcs: first leading from $s_i$ to $s_j$ and the arc leading from $s_i$ to $s_j$ and second leading from $s_j$ to one of its successors $s_{k1}$, ..., $s_{km}$ (Fig. 2) is replaced by single arc leading directly from $s_i$ to the successor of $s_j$ (Fig. 3).

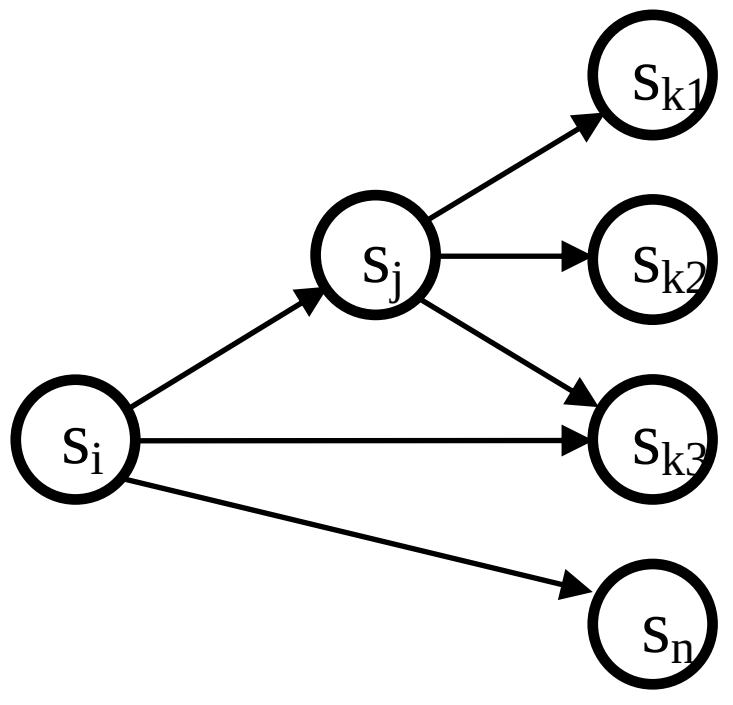

**Fig. 2. A fragment of RG$_{--@}$ before reduction**

Let us define the invisibility of an arc.

**Definition 1.** An arc leading from $s_i$ to its successor $s_j$ is **invisible**, according to a given set of atomic propositions $\pi$, if every proposition $\varphi$ belonging to $\pi$ has the same boolean value in $s_i$ and in $s_j$ ($\varphi$ is *true* in both states or $\varphi$ is *false* in both states): $\pi(s_i)\cap\pi=\pi(s_j)\cap\pi$, $\sim\pi(s_i)\cap\pi=\sim\pi(s_j)\cap\pi$.

## V. Reduction rule

First a reduction of a single state in RG$_{—@}$ will be defined. The principle of reduction is following:

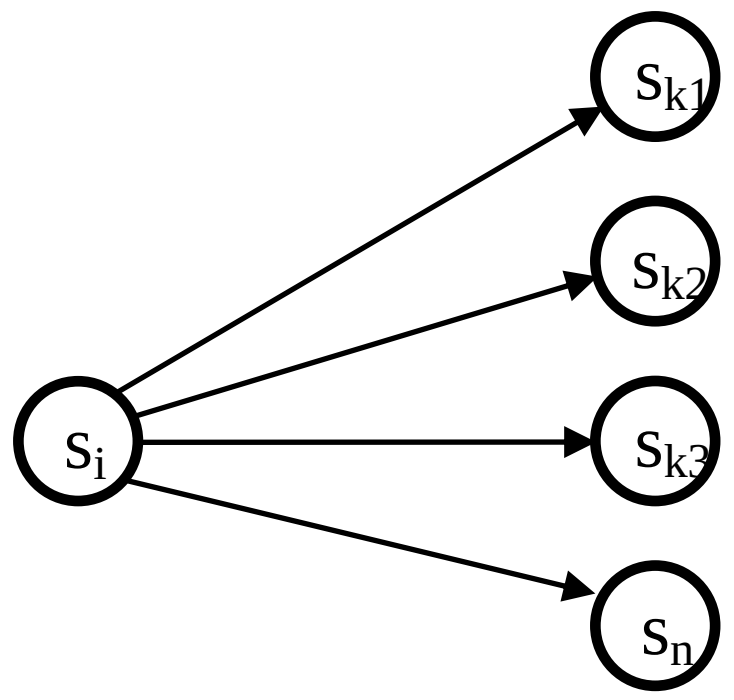

**Fig. 3. A fragment of RG$_{-@}$ after reduction**

**Reduction rule.** In a given RG$_{—@}$, if the arc $x$ is leading from $s_i$ to $s_j$ is invisible according to a given set of atomic propositions $\pi$, and the state $s_j$ has successors $s_{k1}\ldots s_{km}$, with arcs $x_1$ leading from $s_j$ to $s_{k1}$, $x_2$ from $s_j$ to $s_{k2}$, … $x_m$ from $s_j$ to $s_{km}$, then the RG$_{—@}$ may be **reduced** by replacing subset of arcs $x$, $x_1$, …, $x_m$ by subset of arcs $y_1$, …, $y_m$, where $y_1$ leads directly from $s_i$ to $s_{k1}$, $y_2$ from $s_i$ to $s_{k2}$, …, $y_m$ from $s_i$ to $s_{km}$. The reduction may be applied to every arc $x_p$, $p=1..m$ individually, i.e. it is possible that for some $p$ the reduction is applied and for some $n$, $n\neq p$ the reduction is not applied. The state $s_j$ is preserved in RG$_{—@}$ is the latter case.

After reduction, a state $s_j$ disappears from paths going through states $s_{k1}..s_{km}$, for which a reduction was applied. State $s_j$ will be called **skipped** on these paths.

The reduction may be applied under certain conditions:

**Reduction restrictions.**

i. if $s_i$=$s_j$ (an ear), the reduction cannot be applied;
ii. if $s_j$=$s_{k1}$, $m=1$ (an ear), the reduction cannot be applied;
iii. if there is a state $s_{kn}$=$s_i$, then the reduction cannot be applied to the arc $x_n$, therefore the arcs $x$ and $x_n$ are preserved (to prevent creating new ear), yet still other arcs $x_p$, $p=1..m$, $p\neq n$, may be reduced to $y_p$.;
iv. if the proposition is of a form $s_j$:$\varphi$, where the state $s_j$ can be statically found, or $s_j$ belongs to a set $S$ defining a range of a state quantifier (the set $S$ must be statically obtained), then the reduction cannot be applied;
v. there is an exception to restriction *iv*, when all the successors of $s_i$ are also successors of $s_j$ (Fig. 4): in this case reduction may be applied.

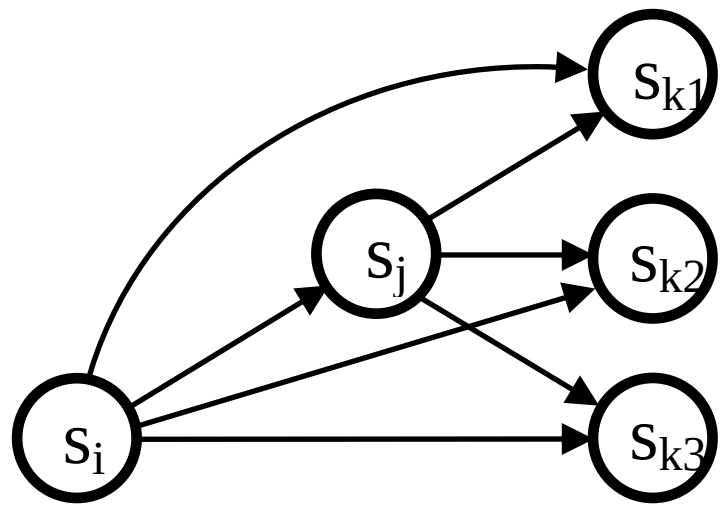

**Fig. 4. Illustration of reduction restriction *v***

**Comment 1.** An example of a range of a quantifier that cannot be statically obtained is following:

$\forall$ s $\in$ FUT(p); s: φ ■

*FUT(p)* denotes a set of states accessible form *p* (being a future of *p* [Dasz, Dasz00a])

**Comment 2.** If a set defining a range of a quantifier cannot be statically obtained, no reduction can be applied (no state can be deleted from $RG_{-@}$).

## VI. Proposition set completeness

**Proposition set completeness.** In some cases, specific propositions must be added to a set $\pi$ to guarantee that all temporal propositions over $\pi$ have the same boolean value in both full $RG_{—@}$ and $RG_{—@}$ with $s_j$ deleted ($RG_{—@-x}$):

i. if the next-step operator ○ is used, it must refer to a state *s* that may be statically found or is a state variable running through a set *S* defining a range of a state quantifier (the set *S* must be statically obtained); in this case a proposition ***in*** *s* must be added to $\pi$ for every such state *s* or each element of set *S*;
ii. the next-step operator ○ may refer only directly to the state (it may not be nested); if the next-step operator in nested, no reduction may be applied;
iii. if the next-in-automaton operator $\circ_{\boldsymbol{a}}$ is used, and it refers to a state *s* that may be statically found or is a state variable running through a set *S* defining a range of a state quantifier (the set defining the range of the quantifier must be statically obtained), in this case a proposition ***in*** *s*|***a*** must be added to $\pi$ for every such state s or each element of set *S*;
iv. if the next-in-automaton operator $\circ_{\boldsymbol{a}}$ is used, and it refers to a state *s* that cannot be statically found or is a state variable running through a set *S* defining a range of a state quantifier, and the set *S* cannot be statically obtained, then a proposition ***in*** $p_{\boldsymbol{a}}$ must be added to $\pi$ for every state $p_{\boldsymbol{a}}$ of the component automaton ***a***.

The set satisfying the above completeness conditions will be called **complete** and denoted $\pi_c$.

## VII. Correctness of the reduction

**Theorem 1.** If in RG$_{_@}$ there is an invisible arc $x$ leading from $s_i$ to $s_j$, and the reduction rule holds for $x$ over $\pi_c$ and reduction restrictions do not have effect, then RG$_{_@}$ with a reduction applied (RG$_{_@\text{-x}}$) may be used as well as RG$_{_@}$ to evaluate all formulas over a $\pi_c$..

To prove the theorem, first we show the stuttering bisimulation between states in RG$_{_@}$ and in RG$_{_@\text{-x}}$.

**Definition 2.** Two states $s$ and $s'$ in Kripke structures $K$ and $K'$, are in **stuttering simulation** relation [Gert99], if for every path $\sigma$ in $K$ starting from $s$ there exists a path $\sigma'$ in $K'$ starting from $s'$, and every of these two paths may be split into partitions (sequences of states) $B_1$, $B_2$, … in $K$, $B_1'$, $B_2'$, … in $K'$, such that for each $h>0$, $B_h$ and $B_h'$ are nonempty and finite, and for every state $s_p$ in $B_h$ and every state $s_p'$ in $B_h'$ $\pi(s_{p1})\cap\pi_c=\pi(s_{p2}')\cap\pi_c$, $\sim\pi(s_{p1})\cap\pi_c=\sim\pi(s_{p2}')\cap\pi_c$. States $s$ and $s'$ will be called **stuttering bisimilar**.

**Proof.** As shown in [Gert99], for any pair of stuttering bisimilar states $s$ and $s'$, $K,s \models \varphi$ iff $K',s' \models \varphi$ ($\varphi$ is a CTL$_{\text{-X}}$ formula). Indeed, every not skipped state $s_p$ in RG$_{_@}$ and in RG$_{_@\text{-x}}$ ($K$ and $K'$) is stuttering bisimilar, even if skipped state $s_j$ is in future of $s_p$ in RG$_{_@}$:

- For every path starting from $s_p$, not going through $s_j$ in RG$_{_@}$, the sequences of states in $K$ and in $K'$ are identical (partitions contain simply single, the same state, Fig. 5).
- For every path starting from $s_p$, going through $s_j$ in RG$_{_@}$ (and therefore through $s_i$):
  - ♦ partitions from $s_p$ to the predecessor of $s_i$ in RG$_{_@}$ and in RG$_{_@\text{-x}}$ contain single state;
  - ♦ partition $B_h$ containing $s_i$ consists of $s_i$ and $s_j$ (as $\pi(s_i)\cap\pi_c=\pi(s_j)\cap\pi_c$), while partition $B_h'$ consists of $s_i'$ only;
  - ♦ next partitions $B_g$ and $B_g'$, $g>h$, contain single state (Fig. 6).

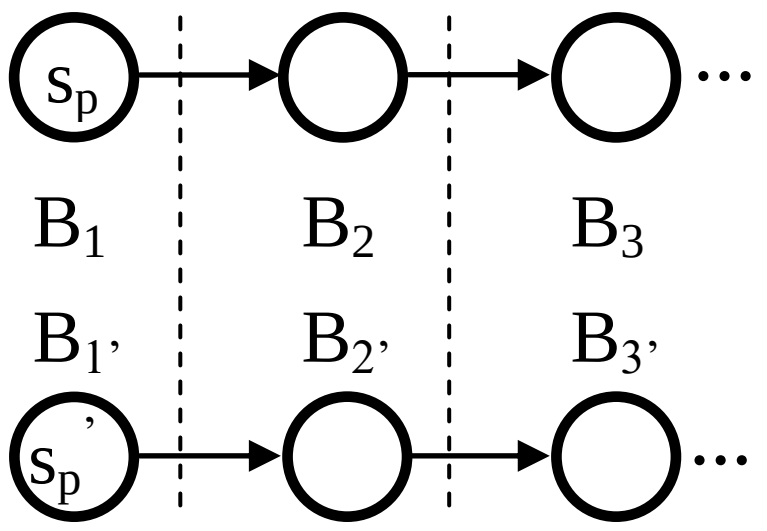

**Fig. 5. Paths not going through $s_j$**

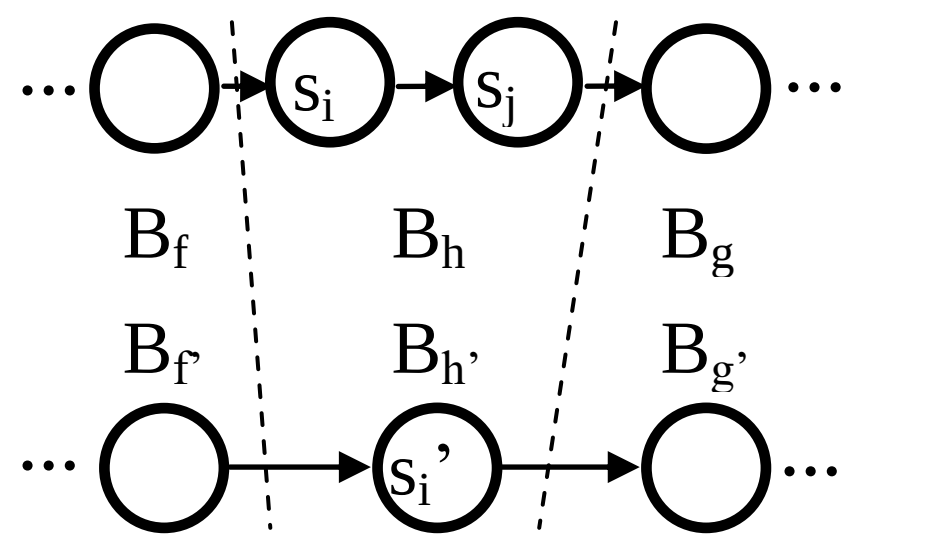

**Fig. 6. Paths going through $s_j$**

Finding stuttering bisimilar paths going through $s_j$ is easy because for every path going through $s_i$, $s_j$ and $s_{kn}$ in RG$_{—@}$ there is exactly one path going through $s_i$ and $s_{kn}$ in RG$_{—@\text{-x}}$.

A state $s_j$ is not stuttering bisimilar with skipped state $s_i$', because if we try to take $s_j$ in RG$_{—@}$ (*K*) and $s_i$*'* in RG$_{—@\text{-x}}$ (*K'*) as stuttering bisimilar:

- we will find a stuttering bisimilar path in RG$_{-@\text{-x}}$ for every path in RG$_{—@}$ (partition $B_1$ consists of $s_j$ only, while partition $B_1$*'* consists of $s_i$*'* only, next partitions contain single, the same state),
- we will not find a stuttering bisimilar path in RG$_{—@}$ for path in RG$_{—@\text{-x}}$ that starts from $s_i$ and has a successor of $s_i$ other than $s_j$ and $s_{k1}$, …, $s_{km}$ as second element (a path *($s_i$, $s_n$, ...)* in Fig. 3; there is no arc from $s_j$ to $s_n$ in RG$_{—@}$ !).

This is the reason why if a formula $s_j$*:*$\varphi$ appears, no state $s_j$ can be skipped (reduction restriction *iv*). Then, the exception (reduction restriction *v*) is following:

- for every path in RG$_{—@}$ starting from $s_j$ there exists a stuttering bisimilar path in RG$_{--@\text{-x}}$ (partition $B_1$ consists of $s_j$ only, while partition $B_1$*'* consists of $s_i$*'* only, next partitions contain single, the same state),
- there are not any successors of $s_i$ other than $s_j$ and $s_{k1}$, …, $s_{km}$, therefore the previous case is the only one (see Fig. 4). ■

## VIII. Reduced reachability graph

The reduction rule lets reduce graphs of a shape "rhombus with diameter" to the diameter only (arc *($s_i$, $s_{k1}$)* in Fig. 7). It is because the CSM model is based on coincidences, not on interleaving. In interleaving models, no "diameter" *($s_i$, $s_{k1}$)* exists in cases like in Fig. 7, and the reduced graph consists of a pair of arcs: either *(($s_i$,$s_{j1}$), ($s_{j1}$,$s_{k1}$))* or *(($s_i$,$s_{j2}$), ($s_{j2}$,$s_{k1}$))*.

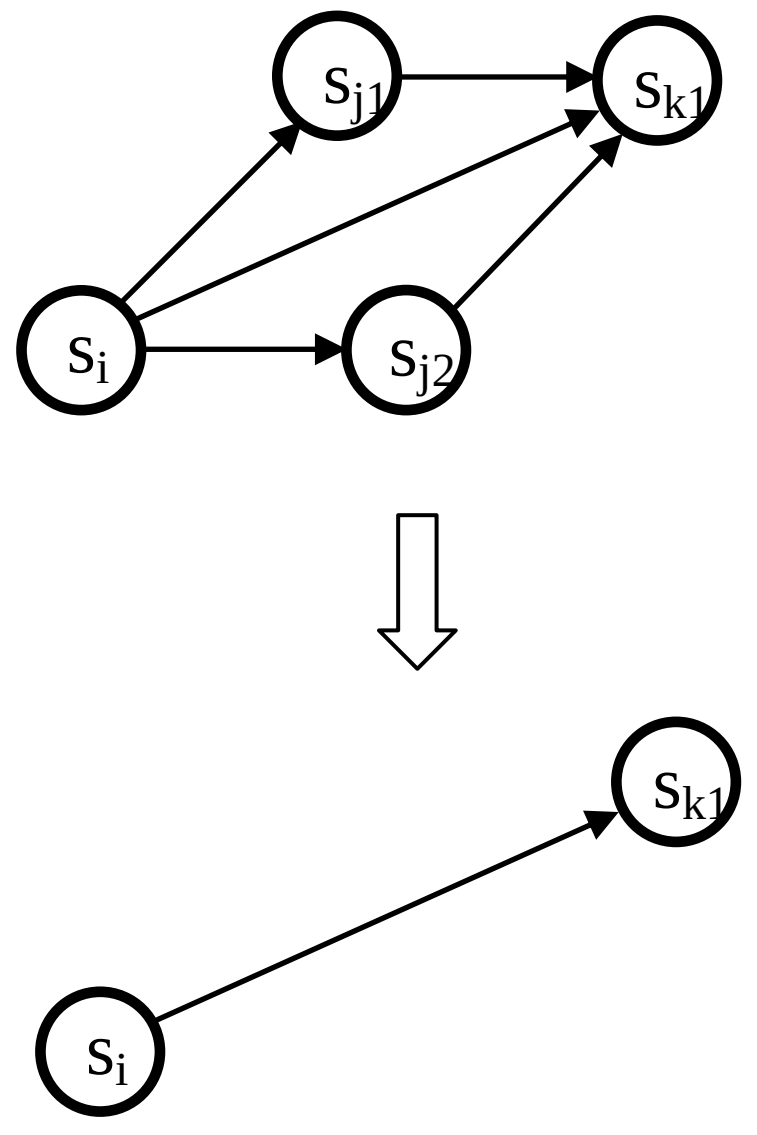

**Fig. 7. Reduction of "rhombus with diameter"**

As the correspondence between states in RG$_{—@}$ and in RG$_{—@\text{-x}}$ is symmetric, the relation is stuttering bisimulation [Gert99]. For every pair of stuttering bisimilar states *s* and *s'* it holds that *RG$_{—@}$,s* $\models \varphi$ iff *RG$_{—@\text{-x}}$,s'* $\models \varphi$ (we showed that *s* cannot be skipped). The next-step operators (○ and ○$_{\mathbf{a}}$) may be used in formulas $\varphi$, because the operators never refer to reduced states or their predecessors (because of atomic formulas of type ***in*** *s* included in $\pi_c$).

After reduction of *x*, next invisible arcs may be reduced from RG$_{—@}$. We will name a **reduced reachability graph** (RRG) a RG$_{—@}$ in which a number of reductions are applied. A **minimal** RRG (MRRG) will be a graph in which no further reduction can be applied. MRRG has not a canonical form – it may depend on an order in which candidates for reduction are taken.

As the next-step operator ○ may be used in formulas over $\pi$ (but states, to which the operator refers, cannot be skipped), and because there is no assumption on the independence of actions, the scope of sets of atomic propositions $\pi$ is much wider than in [Gert99].

The RRG is smaller of reduced arcs ('*x*') and skipped states ('$s_j$').

# IX. The algorithm

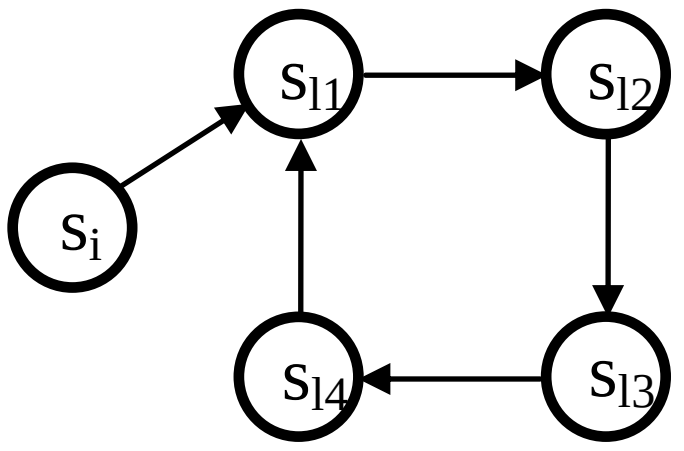

**Fig. 8. A cycle in RG$_{—@}$ subject to reduction**

Basing on the reduction principle presented above, an algorithm of reduction may be proposed. Three additional assumptions should be taken to construct the algorithm:

a) In a situation similar to illustrated in Fig. 8, attempts to apply the reduction may be done infinitely in a cycle in RG$_{—@}$. First $s_{l1}$ is skipped, then $s_{l2}$, $s_{l3}$, $s_{l4}$ and $s_{l1}$ again. The solution is to finish attempts to skip the successors of $s_i$ when a cycle is identified (a state $s_{l1}$ is tried to be skipped for the second time). As a result, $s_{l4}$ will be chosen as a successor of $s_i$ (Fig. 9).

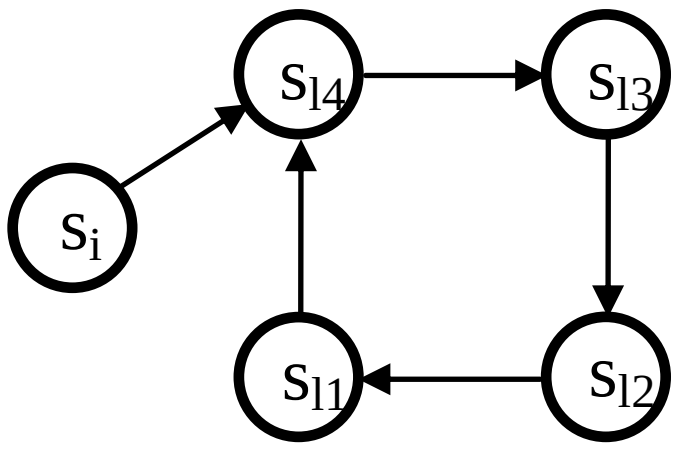

**Fig. 9. The result of reduction with a "trap"**

b) In a similar situation to a previous one (circle like in Fig. 8), a searching for successors in a cycle (using the reduction rule and the above assumption (a)) will fall into a "trap". Having the arc *($s_i$, $s_{l4}$)* after skipping successors of $s_i$, successors of states in a circle are searched. As a result, the following arcs are taken to reduced graph: *($s_{l4}$, $s_{l3}$), ($s_{l3}$, $s_{l2}$), ($s_{l2}$, $s_{l1}$), ($s_{l1}$, $s_{l4}$)*. Surprisingly, no reduction is achieved (although the graph has different shape). The result is illustrated in Fig. 9. This weakness of the reduction can be solved by a principle that if a state already taken to the resulting graph is tried to be skipped, the algorithm stops searching for next successors. Using this assumption, a graph shown in Fig. 10 will be achieved.

c) A "one-pass" algorithm does not guarantee that the reduced graph is minimal (it is RRG, not necessary MRRG). It is possible that starting from initial state of reduced graph, new reductions are possible. However, the assumption is made that the "one-pass" reduction is enough.

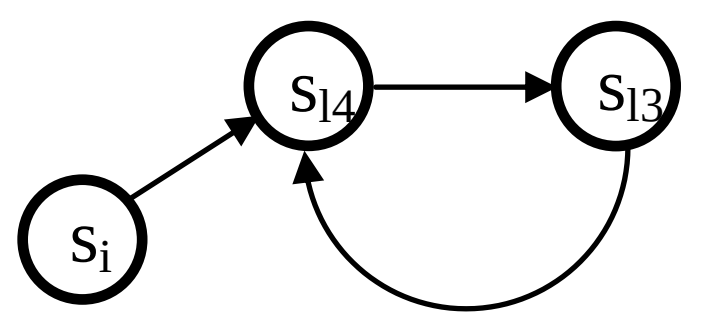

**Fig. 10. The result of reduction without a "trap"**

The first version of the algorithm works on fully obtained $RG_{\neg @}$ (off-line version):

1. Mark the initial state as preserved.
2. Mark all states as unchecked.
3. Get one of preserved and unchecked states.
   If all preserved states are checked, stop the algorithm.
4. Find the successors of the state taken in p.3.
5. Check which successors found in p.4 (or p.6) may be skipped.
   Additional constraint: if a state under check was already checked in analysis of the state taken in p.3, it cannot be skipped (this prevents infinite analysis of successors forming a circle).
   If nothing may be reduced, go to 8.
6. Find the successors of states reduced in p.5.
7. Go to 5.
8. Mark the state taken in p.3 as checked, and not skipped successors (or successors of successors, …) as preserved.
9. Go to 3.

The on-line version of the algorithm is performed during elaboration of $RG_{\neg @}$.

1. Obtain the initial state, mark it as preserved.
2. Mark the initial state as unchecked.
3. Get one of preserved and unchecked states.
   If all preserved states are checked, stop the algorithm.
4. Obtain the successors of the state taken in p.3.
5. Check which successors obtained in p.4 (or p.6) may be skipped.
   Additional constraint: if a state under check was already checked in analysis of the state taken in p.3, it cannot be skipped (this prevents infinite analysis of successors forming a circle).
   If nothing may be reduced, go to 8.
6. Obtain the successors of states reduced in p.5.
7. Go to 5.
8. Mark the state taken in p.3 as checked, and not reduced successors (or successors of successors, …) as preserved.
9. Go to 3.

While running this reduction algorithm, the reduction restriction *v* may be even relaxed: if more than one successor of a state are candidates to be skipped, and some of them are used in formulas $s{:}\varphi$, and every successor of $s_i$ is also a successors of at lest one candidate to be skipped, then all candidates may be reduced at once.

## X. Time complexity

The time complexity of the algorithm is:

- For off-line version, for every state at most all arcs (successors, successors of successors, etc.) should be checked, therefore the complexity is $O(N^3)$ where N is the number of states in RG (the number of arcs is at most $N^2$).
- For on-line version, every state must be taken to obtain its successors. For every taken state, at most all arcs (successors, successors of successors, …) should be checked, therefore the complexity of obtaining RRG is multiplied by $O(N^2)$, where N is the number of states in full RG.

The worst-case complexity is poor, but it is achieved in case of large number of arcs in $RG_{—@}$ (a graph near to a clique). Yet the reduction in such graphs may be substantial. In more "linear" graphs reduction is less (only invisible sequences generally) and the time of reduction is near to O(N).

## XI. Conclusions

Although no independence of actions may be found in CSM systems, and the principle of semantics of CSM is coincidence of actions rather than interleaving, a reduction rule was found and an efficient reduction algorithm was proposed. The weak point of the rule is that all successors of a state and successors of successors (two arcs depth) must be evaluated before reduction (while the algorithm proposed in [Gert99] may choose preserved arcs not evaluating all arcs in some cases). Therefore, not savings may be acquired in time of obtaining reduced RG. But saving in space required for storing state space may be substantial.

The advantage of proposed reduction rule is that it may be used even when a next-step operator ○ is used. Also, because of natural "diameters" in state space (coincidences), the reduction ratio (number of states and arcs in RG/ number of states and arcs in $RG_{—@}$) may be greater than for interleaving systems.